\documentclass[prb,notitlepage,showpacs,preprintnumbers]{revtex4-2}

\usepackage[utf8]{inputenc} 
\usepackage{graphicx}
\usepackage{ifthen}
\usepackage{ifpdf}
\usepackage{color}
\definecolor{red}{rgb}{1.,0.,0.}
\definecolor{blue}{rgb}{0.,0.,1.}

\usepackage{latexsym}
\usepackage{amsmath}
\usepackage{bm}
\usepackage{bbm} 
\newcommand{\half}{{\textstyle\frac{1}{2}}}

\newcommand{\re}{\mbox{Re}}
\newcommand{\eexp}{\mbox{e}^}

\newcommand{\beq}[1]{\begin{eqnarray}\ifthenelse{#1=-1}{\nonumber}
		{\ifthenelse{#1=0}{}{\label{e#1}}}}
	\newcommand{\eeq}{\end{eqnarray}}
\newcommand{\be}{\begin{equation}}
\newcommand{\ee}{\end{equation}}

\newcommand{\bea}{\begin{eqnarray}}
\newcommand{\eea}{\end{eqnarray}}

\newcommand{\hide}[1]{}

\newcommand{\tr}{\mathop{\rm Tr}}

\makeatother
\makeatletter

\begin{document}

\title{Theory of single molecule NMR detected by real time ESR   \\ \vspace{.0cm}}
\author{Baruch Horovitz$^{1}$ and Alexander Shnirman$^{2,3}$ }
\affiliation{$^1$Department of Physics, Ben Gurion University of the Negev, Beer Sheva 84105, Israel\\
	$^2$Institute for Theoretical Condensed Matter Physics, Karlsruhe Institute of Technology, 76131 Karlsruhe, Germany\\
	$^3$Institute for Quantum Materials and Technologies, Karlsruhe Institute of Technology, 76344 Eggenstein-Leopoldshafen, Germany}
\begin{abstract}
	In relation to recent experimental data [Y. Manassen et al., J. Mag, Res. {\bf 374}, 107863 (2025)] , we develop a theory framework for demonstrating the feasibility of detecting sharp Nuclear Magnetic Resonance (NMR) oscillations in a real time ESR data. The procedure is to follow real time oscillations of the ESR signal measured at a selected frequency of a hyperfine transition. We study a variety of systems such as a single radical molecule with one or two hyperfine coupled nuclei or two molecules that coexist as either radicals or non-radicals, facilitated by charge transfer. We develop a master equation for describing these scenarios and find parameters for which a sharp NMR line can be observed. We show that in all cases an off-diagonal term in the hyperfine tensor is essential for the observation.
\end{abstract}
\maketitle

The detection of Nuclear Magnetic Resonance (NMR) of individual molecules is an outstanding challenge. This would enable methods in chemical analysis, quantum information, and medical NMR. A variety of methods were attempted: the recent real-time ESR method \cite{manassen}, magnetic resonance force microscopy \cite{mamin,degen}, using NV centers \cite{muller,pfender,gulka}, optically detected magnetic resonance \cite{wrachtrup,gao}, rf spin pumping \cite{stolte}, transport methods \cite{vincent,du} or STM-ENDOR methods \cite{manassen3,vennema}. Recently a technique was developed for a real time analysis of electron-spin-resonance by scanning-tunneling-microscope (ESR-STM) signals, that via a hyperfine coupling can detect NMR oscillations \cite{manassen}. 

In this paper we develop a theory for NMR detection via real time ESR in a single molecule. The input data is single molecule ESR intensity as function of time, at a chosen frequency in the vicinity of a hyperfine line. The ESR data can be detected by either an STM current noise \cite{manassen} or by other methods \cite{baumann,willke,kovarik}. Our theory is independent of how the ESR is measured, however, it is essential that this input data is measured during time steps that are shorter than the nuclear precession period as well as longer than the electron precession time, as achieved in Ref. \onlinecite{manassen}; this scheme is effectively a non-equilibrium detection.  These conditions are needed so that the ESR is well defined (i.e. many ESR precessions) and yet the nuclear spin is at an instantaneous value within its precession.
In the following, whenever we mention NMR we mean this particular single molecule NMR detected by real time ESR.

 Our real time method is therefore distinct from all previous methods for single molecule NMR \cite{mamin,degen,muller,pfender,gulka,wrachtrup,gao,stolte,vincent,du,manassen3,
vennema}. Furthermore, we note that our method yields the NMR spectra of nuclei in a single molecule to an accuracy of a few Hz with a linewidth of $\sim$ 10 Hz \cite{manassen}, allowing e.g. for detection of chemical shifts. Combining the STM scanning capability and its spatial resolution of $\approx$ 1 nm show that this method is superior to the previous methods \cite{mamin,degen,muller,pfender,gulka,wrachtrup,gao,stolte,vincent,du,manassen3,
	vennema}.

We find a few scenarios depending on parameters that result in sharp NMR lines. In all cases an off-diagonal hyperfine element is essential, as possible in a rotated molecule relative to the magnetic field axis (see Appendix A). We first consider a single radical molecule that has a nucleus with a weak hyperfine couplings relative to the ESR linewidth. The resulting sharp NMR line is a manifestation of motional narrowing. A variant of this case is a radical with two nuclei, one that has weak hyperfine coupling showing sharp NMR, while the second nucleus has strong hyperfine coupling, allowing for hyperfine splitting of the ESR line. 

An extension of this model allows the presence of a second molecule allowing charge transfer between the molecules (ionization), so that both molecules coexist as either radicals or non-radicals. A second radical is actually essential for observing ESR-STM \cite{bh}. We consider strong intermolecular dissipative transitions while the intra-molecular electron spin relaxation is weak in one molecule and strong in the other one.
Within our master equation we find scenarios, in which a sharp NMR resonance is possible
while the ESR spectrum is broadened and hyperfine splitting is not seen.
We have extended the model to a non-equilibrium situation, where the rates of forward and backward ionization transitions are distinct and alternate in time. This alternation is caused by the voltage modulation as in the experimental data \cite{manassen}. Unlike the equilibrium case we find that one can observe NMR and simultaneously hyperfine split ESR.
To determine which scenario corresponds to the actual data a detailed ESR study is needed. A brief summary of all cases is provided in table I in the conclusions.

\section{Single molecule case}

\subsection{NMR with one nuclear spin}

We consider a single electron spin coupled to a single nuclear spin. The Hamiltonian reads 
\beq{01} 
{\cal H}=\half\nu_e\sigma_z +\half \nu_n \tau_z+\sigma_z\left(a\tau_{z} +d\tau_{y}\right) 
=\half\nu_e\sigma_z +\half \nu_n \tau_z + \tilde a \sigma_{z} \tilde \tau \ ,
\label{single}
\eeq
where $\nu_e, \,\nu_n$ are the electron and nuclear free precession frequencies, respectively, and ${\bm\sigma},\,{\bm\tau}$ are Pauli matrices for the electron and nuclear spins, respectively. A rotated molecule has in general an off-diagonal term, i.e. $d\neq 0$, see Appendix A. We neglect other hyperfine elements, e.g. $\sim \sigma_x\tau_x$, as these lead to second order shifts.

We have introduced 
\beq{02}
\tilde \tau \equiv \frac{1}{\tilde a}\left(a\tau_{z} + d\tau_{y}\right)\quad,\quad \tilde a = \sqrt{a^{2}+d^{2}}\ .
\eeq 

We assume that only the electronic spin is subject to relaxation (either due to the coupling to the thermal bath or 
due to the continuous ESR measurement). This process is characterised by two rates: $\Gamma_{d}$ - 
the rate of spin flips $\uparrow \rightarrow \downarrow$ and  $\Gamma_{u}$ - 
the rate of spin flips $\downarrow \rightarrow \uparrow$.

The simplest scenario is when 
the electronic spin polarization $\sigma_z$ fluctuates on a time scale much faster than $\nu_n$ and averages almost to 
zero. This scenario, akin to the motional narrowing, allows for a sharp NMR resonance. It requires $\Gamma_1\gg \tilde a$, where $\Gamma_1\equiv \Gamma_{u}+\Gamma_{d}$ is the longitudinal relaxation rate of the electronic spin (related to the ESR line width).
More specifically, the time evolution of the nuclear raising operator $\tilde\tau_+$ (that describes the nuclear spin precession), neglecting here $\nu_n$, is given by
\beq{03}
\tilde\tau_+(t)=\eexp{-i\tilde a \tilde\tau\int_0^t \sigma_z(t')dt'}\tilde\tau_+\eexp{i\tilde a \tilde\tau\int_0^t\sigma_z(t')dt'}=\eexp{-2i\tilde a \int_0^t\sigma_z(t')dt'}\tilde\tau_+\ .
\eeq
To average over the fast dynamics of $\sigma_{z}(t)$ we use
\beq{03}
\langle \delta \sigma_{z}(t) \delta\sigma_{z}\rangle_{\omega} = (1-\langle \sigma_{z}\rangle^{2})\frac{2\Gamma_{1}}{\omega^{2}+\Gamma_{1}^{2}}\ ,
\eeq
where $\delta\sigma_{z}\equiv \sigma_{z} - \langle \sigma_{z}\rangle$ and $\langle \sigma_{z}\rangle=(\Gamma_{u}-\Gamma_{d})/\Gamma_{1}$.
As a result one obtains
\beq{03}
\langle \tilde\tau_+(t)\rangle\sim \langle\eexp{-2i\tilde a \int_0^t\sigma_z(t')dt'}\rangle=
\eexp{-2i\tilde a \langle\sigma_{z}\rangle t}\,\,\eexp{-\frac{4\tilde a^2}{\Gamma_1}(1-\langle \sigma_{z}\rangle^{2})  t}
\label{gaussian}
\eeq
The first (oscillating) exponent corresponds to a chemical shift of the nuclear Larmor frequency, whereas the second 
(decaying) exponent gives the nuclear line width.
If $\nu_n$ is taken into account, the coupling to the nuclear spin is not purely longitudinal, i.e., $d\neq 0$ and $\tilde \tau \neq \tau_{z}$, and the dephasing rate of the nuclear spin is small, i.e., $4(1-\langle \sigma_{z}\rangle^{2})\tilde a^2/\Gamma_1\ll \nu_n$ (motional narrowing), we expect $\tilde \tau$ to perform slowly decaying oscillations with 
frequency close to $\nu_n$, as needed for observing NMR. 
This derivation is extended in Appendix B to the case of small $\Gamma_1$. 

We write down the Lindblad equation governing the dynamics of the system's density matrix 
\beq{077}\label{eq:Lindblad}
\frac{d\rho}{dt}=L\rho = -i[{\cal H},\rho]+
\Gamma_{d}\left[\sigma_{-}\,\rho\,\sigma_{+} - \half \sigma_{+}\sigma_{-} \rho - \half \rho \sigma_{+}\sigma_{-} \right ] +
\Gamma_{u}\left[\sigma_{+}\,\rho\,\sigma_{-} - \half \sigma_{-}\sigma_{+} \rho - \half \rho \sigma_{-}\sigma_{+} \right ] \ .
\eeq
 In general there is also a pure dephasing rate $\Gamma_\varphi$ that represents fluctuations coupled to $\sigma_z$ and then the ESR linewidth is $\Gamma_2=\half\Gamma_1+\Gamma_\varphi$ . It was shown that $\Gamma_1$ and $\Gamma_\varphi$ are comparable in the STM scenario with high voltage \cite{bh,paaske}. We neglect here $\Gamma_\varphi$ to emphasize the spin relaxation effect, otherwise $\Gamma_\varphi$ would effectively enhance $\Gamma_2$.
The Lindbladian Eq. \eqref{e077} allows investigating the ESR response function $C^{-+}(\omega)$: 
\bea
C^{-+}(\omega)=\int_0^\infty\langle\sigma_-(t)\sigma_+(0)\rangle\eexp{i\omega t} dt\ .
\eea
We will be interested (for $\omega>0$) in 
\bea
C(\omega)= \int_{-\infty}^\infty\langle\sigma_-(t)\sigma_+(0)\rangle\eexp{i\omega t} dt = 2{\rm Re}\left[C^{-+}(\omega)\right] 
\eea

It is easy to obtain $C(\omega)$ in the limit $\nu_{n}=0$, i.e., when $\tilde\tau$ is a constant of motion.
We obtain
\be
C^{-+}(\omega)=\frac{1-\langle \sigma_{z}\rangle}{2}\,\frac{1}{i(\nu_{e} + 2\tilde a \tilde \tau -\omega) + \Gamma_{2}}\ ,
\ee
which gives
\be
C(\omega) = \frac{1-\langle \sigma_{z}\rangle}{2}\,\frac{2\Gamma_{2}}{(\omega-\nu_{e}-2\tilde a \tilde \tau)^{2}+\Gamma_{2}^{2}}\ .
\ee
We can expand, assuming $2\tilde a \ll \Gamma_{2}$,
\be\label{eq:Cexpandedintau}
C(\omega)\approx \frac{1-\langle \sigma_{z}\rangle}{2}\left[\frac{2\Gamma_{2}}{(\omega-\nu_{e})^{2}+\Gamma_{2}^{2}} 
+ \frac{4\Gamma_{2}\tilde a \tilde \tau\,(\omega-\nu_{e})}{((\omega-\nu_{e})^{2}+\Gamma_{2}^{2})^{2}} \right]\ .
\ee
The correction proportional to $\tilde \tau$ is essentially the overlap of $\tilde \tau$ and $C(\omega)$. It is maximal around $|\omega-\nu_{e}| \approx \Gamma_{1}$. Thus, at its maximum, 
the correction is of relative order $\tilde a/\Gamma_{1}$. This overlap is quantified in the following numerical study.
Since for sharp NMR we need $\tilde a/\Gamma_1\ll\nu_n/\tilde a$, we infer that the ratio of the overlap to ESR signal is bounded by $\nu_n/\tilde a$.

In the opposite limit $2\tilde a \gg \Gamma_{1}$ we obtain two resolved peaks at $\omega = \nu_{e} \pm 2\tilde a$, 
corresponding to $\tilde \tau = \pm 1$. The overlap is then large.

For $\nu_{n} \neq 0$ we resort to numerics. We introduce first super-operators so that objects like $A \rho B$ are written as
\be
(A \rho B)_{ij} = A_{in} \rho_{nm} B_{mj} = A_{in} (B^{T})_{jm}\rho_{nm} = (A \otimes B^{T})_{ij,nm}\rho_{nm}\ .
\label{super}
\ee 
This way $\rho_{nm}$ becomes vectorised with $nm$ being a composite index. Note, that $\tr$, e.g., 
in (\ref{eq:QRegression}) is still the usual trace, namely $\tr(A\rho B)=(A\rho B)_{ii}=(A \otimes B^{T})_{ii,nm}\rho_{nm}$.
We then employ the quantum regression theorem, which gives
\bea\label{eq:QRegression}
\langle\sigma_-(t)\sigma_+(0)\rangle = \tr\left[(\sigma_-\otimes I_d) \, e^{Lt} \, (\sigma_+\otimes I_d) \rho_{\infty}\right]\ .
\eea
Here $\rho_{\infty}$ is the stationary solution of the Lindblad equation $L\rho_{\infty}=0$ where $L$ is the super-operator representation of Eq. \eqref{eq:Lindblad} and $\sigma_{\pm}$ is transformed to super-operators $\sigma_{\pm}\otimes I_d$ as well, $I_d$ is a unit matrix of 
dimension $d=4$ in this case.

Performing the Fourier (Laplace) transform we get 
\beq{10}
C^{-+}(\omega)=\tr[(\sigma_- \otimes I_d)\frac{1}{-i\omega-L}(\sigma_+\otimes I_d)\rho_{\infty}]\ .
\eeq

To proceed with our goal of simulating the NMR-STM experiment, consider
first a qualitative argument. Assume the hyper-fine splitting 
is well resolved, i.e., the ESR spectrum consists of two narrow peaks 
at $\omega_{\pm} = \nu_{e} \pm 2 \tilde a$. This regime is achieved if $\Gamma_{d},\Gamma_{u} \ll \tilde a$.
The peak at $\omega_{+}$ is observed if the nuclear spin is in the state such that $\tilde\tau = 1$, whereas the peak at $\omega_{-}$ is observed if $\tilde \tau = -1$. If the expectation value of $\tilde \tau$ would oscillate slowly 
with frequency close to $\nu_{n}$ the relative weight of the ESR peaks would oscillate accordingly and this could be observed 
if one measures $C(\omega)$ many times near a certain value of $\omega$ (with some bandwidth) and the results are Fourier transformed in order to obtain the power spectrum of the NMR signal. 

Unfortunately, in the considered system with one nuclear spin the scenario indicated above is impossible. 
Indeed, if the hyper-fine peaks are resolved, i.e., $\Gamma_{d},\Gamma_{u} \ll \tilde a$, and $\Gamma_{d},\Gamma_{u}\gg \nu_{n}$, the nuclear spin dynamics is suppressed. In the extreme case $\Gamma_{d},\Gamma_{u} \ll \nu_{n} \ll \tilde a$ the NMR peaks 
reappear at shifted positions (see below).

Interestingly, one can observe the NMR spectrum in the ESR signal in the opposite regime, $\Gamma_{d},\Gamma_{u} >> \tilde a$, 
i.e., when the hyper-fine spectrum is not resolved. More precisely this happens if 
$\frac{\tilde a^{2}}{\Gamma_{u}+\Gamma_{d}}\ll \nu_n$, as in Eq. \eqref{gaussian}. In this regime the effective hyper-fine field experienced by the 
nuclear spin $\tilde a \sigma_{z}$ shows fast fluctuations and averages out. The situation is well known under the name 
''motional narrowing''.

\begin{figure}[t]
	\includegraphics*[height=.27\columnwidth]{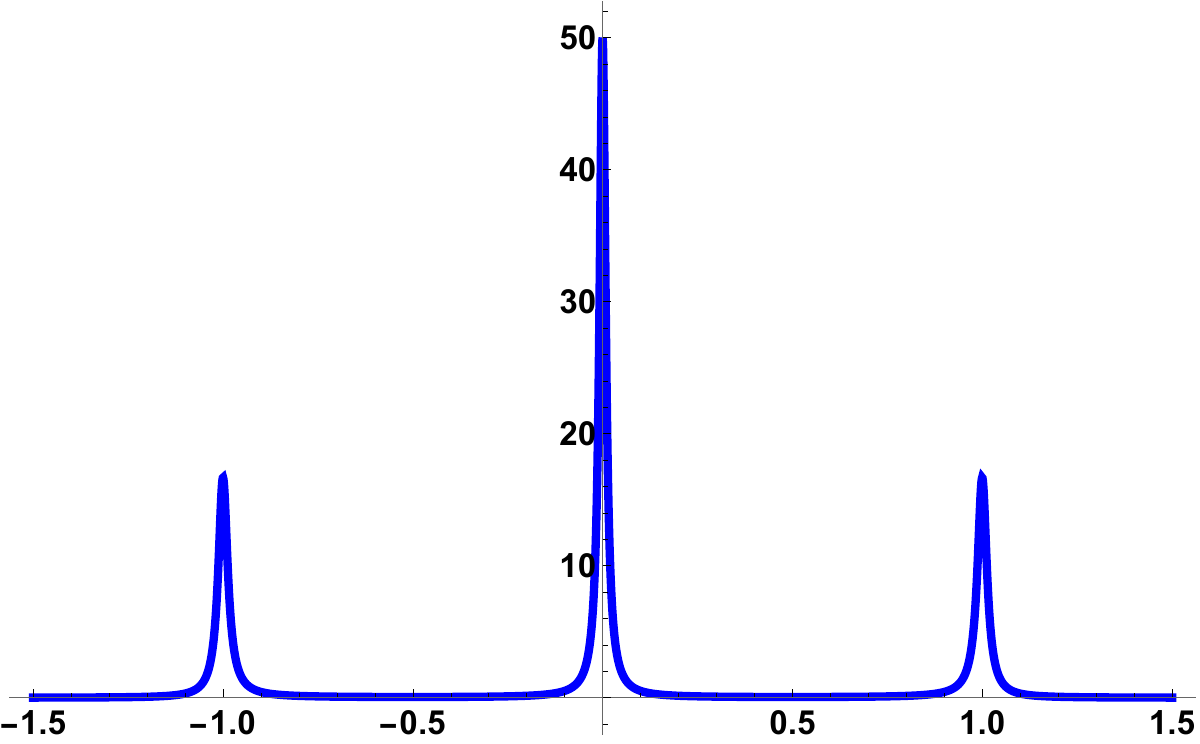}
	\includegraphics*[height=.27\columnwidth]{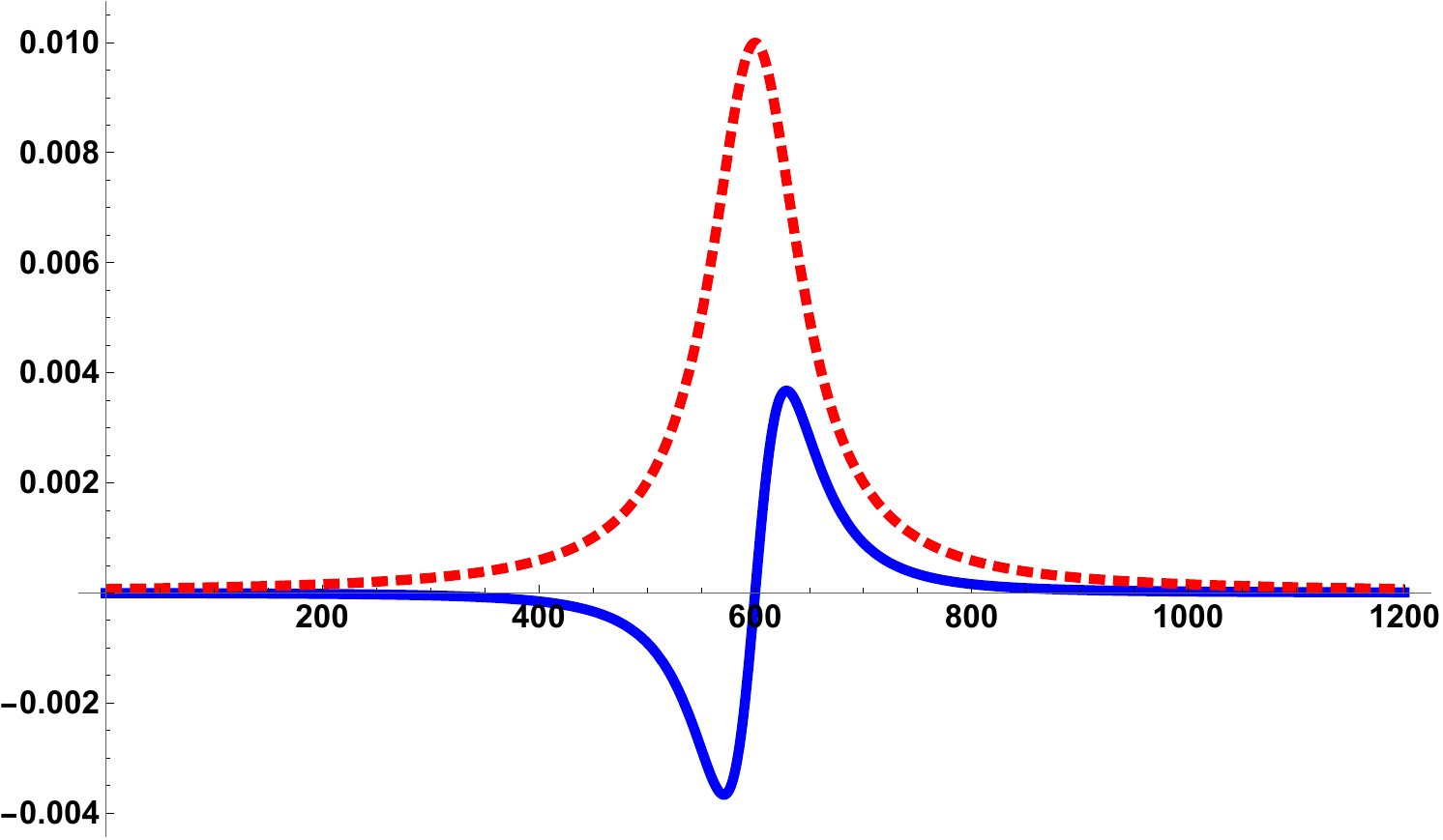}
	\caption{{\bf Left panel:} NMR Spectrum, $C_{nmr}(\nu)$. {\bf Right panel:} 
		The overlap function $\langle \hat C(\omega) \tilde \tau \rangle$ (blue, enhanced by factor 20 for clarity) and the ESR spectrum $C(\omega)$ (red, dashed) as functions of $\omega$. Parameters are $\nu_e=600,\,\nu_n=1,\,a=0.5,\,d=0.5,\,\Gamma_{u}=\Gamma_{d}=50$. We observe a weak overlap of order of $\tilde a/\Gamma_{1}$ .}
	\label{ESRoverlap}
\end{figure}

\begin{figure}[b]
	\includegraphics*[height=.2\columnwidth]{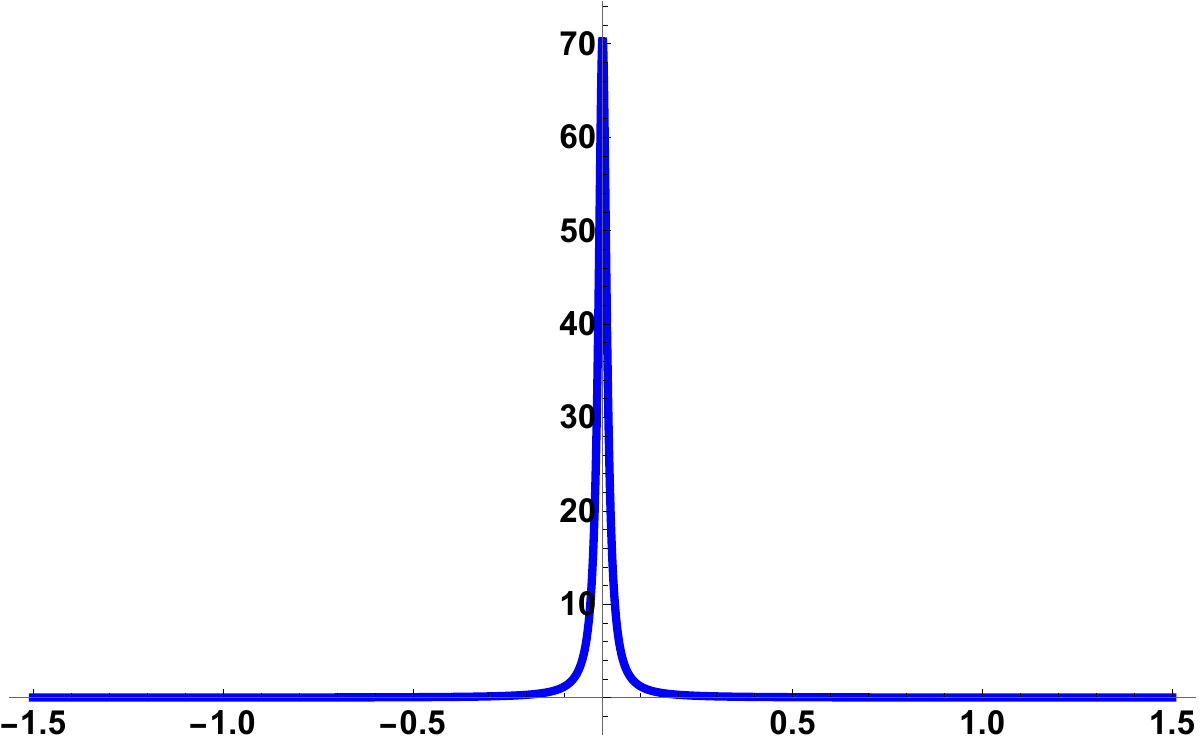}
	\includegraphics*[height=.2\columnwidth]{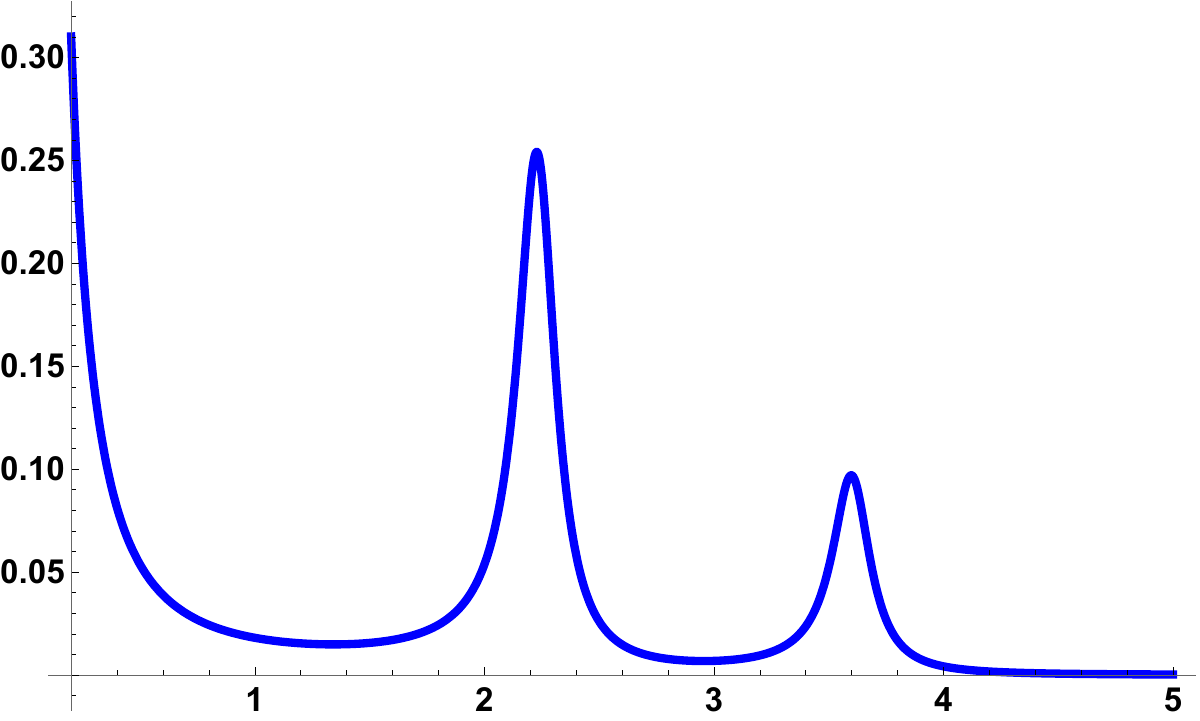}
	\includegraphics*[height=.2\columnwidth]{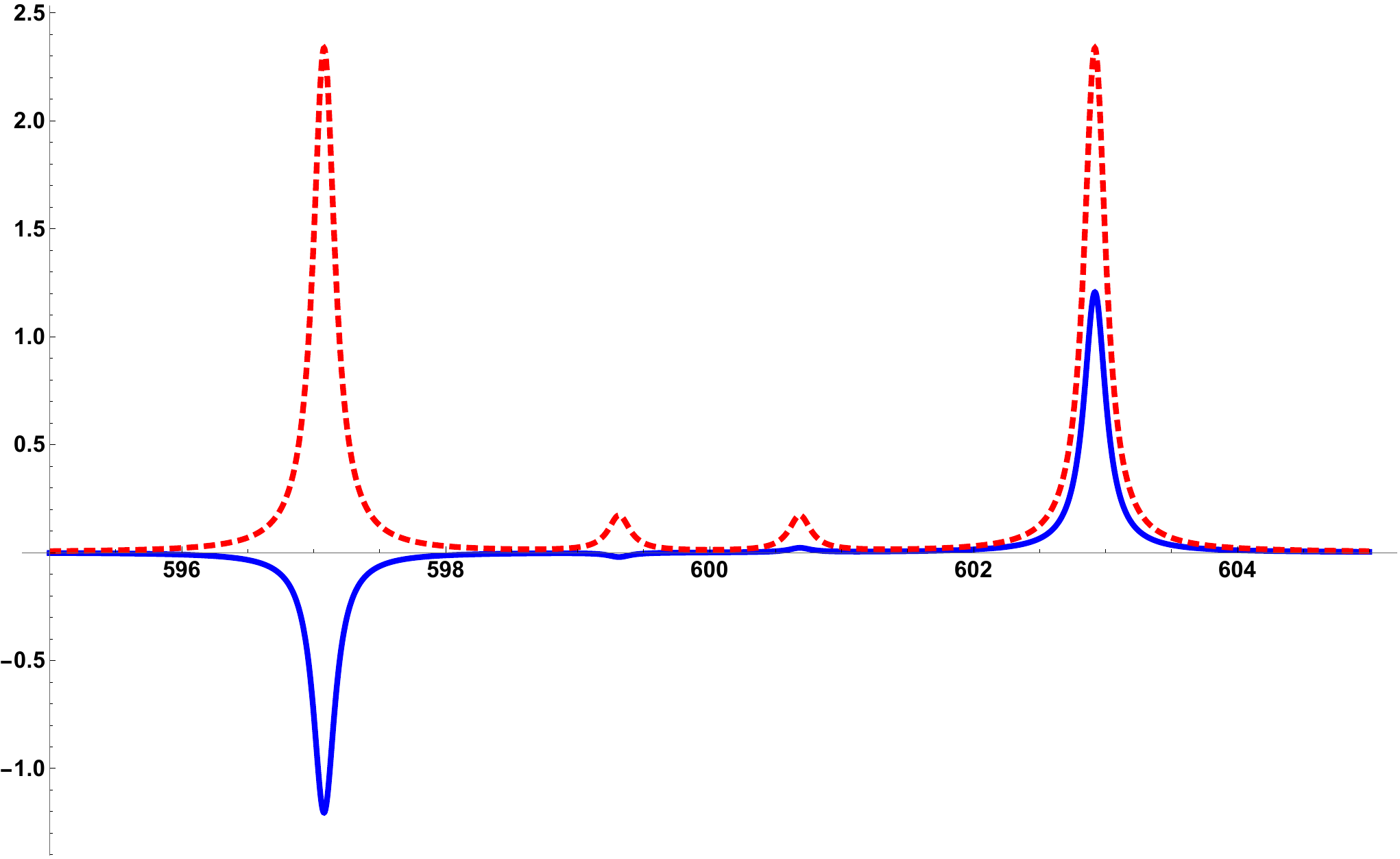}
	\caption{{\bf Left panel:} NMR Spectrum, $C_{nmr}(\nu)$. {\bf Midle panel:} NMR Spectrum, $C_{nmr}(\nu)$, at higher 
		frequencies. The peaks correspond to the hyperfine splittings. {\bf Right panel:} 
		The overlap function $\langle \hat C(\omega) \tilde \tau \rangle$ (blue, reduced by factor 2 for clarity) and the ESR spectrum $C(\omega)$ (red, dashed) as functions of $\omega$. Parameters are $\nu_e=600,\,\nu_n=1,\,a=1.0,\,d=1.0,\,\Gamma_{u}=\Gamma_{d}=0.1$. The overlap between the ESR signal and $\tilde \tau$ is perfect.}
	\label{ESRoverlapresolved}
\end{figure}

We now turn to the main subject of this work, namely the possibility of the ESR signal $C(\omega)$ to show slow NMR 
related oscillations. We assume that the ESR signal is measured in consecutive time bins 
$[t_{N}-\Delta t/2,t_{N}+\Delta t/2]$. Thus the measured signal is $C(\omega, \Delta\omega, t_{N})$, where 
$\Delta \omega \sim 1/\Delta t$ is the band width. The measured signal is spectrally analysed as a function of 
$t_{N}$ at frequency $\nu \ll \Delta \omega$. Effectively, what is measured is the Fourier transform $S(\nu)$ 
of the following correlation function
$S(t_{N_{1}}-t_{N_{2}})=\langle  C(\omega,\Delta\omega, t_{N_{1}})  C(\omega,\Delta\omega, t_{N_{2}})\rangle$.

To estimate the NMR contribution to $S(\nu)$
we 
introduce the following super-operator that represents the ESR absorption spectrum
\be
\hat C(\omega) \equiv  (\sigma_-\otimes I_d) \,\frac{1}{-i\omega - L}\,(\sigma_+\otimes I_d) + h.c.
\label{C0}
\ee
and calculate the overlap between this ESR operator and $\bar\tau$,
\beq{140}
\langle \hat C(\omega) \tilde \tau \rangle = {\rm Tr} \left[\hat C(\omega) \,(\tilde \tau \otimes I_d)\,\rho_{\infty}\right]\ .
\label{nuclearESR}
\eeq

We consider in these expressions $\hat C(\omega)$ as implicit function of the times $t_N$ via its dependence on $\tilde\tau$. We then expand this operator, similar to Eq. \eqref{eq:Cexpandedintau}, as
\beq{71}
\hat C(\omega)=\hat C_0(\omega)+\hat C_1(\omega)\tilde\tau(t)
\eeq
so that $\langle\hat C_1(\omega)\rangle=\langle\hat C(\omega)\tilde\tau\rangle$ is the overlap. Thus the observed ESR slow modulation is
\beq{72}
S(\nu)=\langle\hat C_1(\omega)\rangle^2 \langle\tilde\tau(t)\tilde\tau(t')\rangle_\nu
\eeq
with $\nu$ the Fourier of $t-t'$ that should detect the NMR frequency.
In the numerical solution below we check that $\langle\hat C_1(\omega)\rangle < \langle\hat C(\omega)\rangle$ to justify the expansion. Thus the NMR spectrum
$C_{nmr}(\nu)=<\tilde\tau(t)\tilde\tau(t')>_\nu$ is contained in the slow fluctuations of the ESR signal $C(\omega)$, though we should keep in mind that its overall intensity has a possibly small pre-factor $\langle\hat C(\omega)\tilde\tau\rangle^{2}$. The ESR frequency $\omega$ can be chosen to maximize the overlap.

The NMR spectrum is calculated using the quantum regression theorem, i.e., 
\beq{141}
C_{nmr}(\nu)={\rm Re}\left[\tr[(\tilde\tau\otimes I_d)\frac{1}{-i\omega-L}(\tilde\tau\otimes I_d)\rho_{\infty}]\right]\ .
\eeq

The results are shown in Fig.~\ref{ESRoverlap} with the nuclear correlation function \eqref{e141} in the left panel and the overlap \eqref{e140} in the right panel. 
We note that the nucleus feels an additional effective magnetic field $\tilde a\langle \sigma_z\rangle$ if the electron spin is polarized. However, under an STM bias $V$ the electron spin occupation are determined \cite{bh} by $eV$ rather than temperature $T$, since $eV\gg k_BT$.  Hence we assume $\Gamma_{u}=\Gamma_{d}$ and the spin polarization $\langle\sigma_z\rangle$ vanishes.
The overlap $\langle \hat C(\omega) \tilde \tau \rangle$ can be estimated using Eq.~(\ref{eq:Cexpandedintau}), indeed we observe that the maximum $\langle \hat C(\omega) \tilde \tau \rangle \sim \tilde a/\Gamma_{2}$ is achieved 
at $|\omega - \nu_{e}| \sim \Gamma_{2}$.

For comparison, in Fig.~\ref{ESRoverlapresolved}, we show the results in the limit of well resolved ESR hyperfine lines and $\Gamma_d,\,\Gamma_u\ll\nu_n$. 
In this case we 
see the NMR line at $\nu_{n}$ is completely suppressed. Instead there are weak lines at the shifted positions
$\tilde \nu_{n} \equiv \sqrt{(\nu_{n}\pm
	2 a)^{2}+ 4 d^{2}}$, corresponding to an effective magnetic field seen by the nucleus at either $\sigma_z=\pm$.
Despite the weakness of the spectral lines, the overlap in this case is strong.

\subsection{NMR with two nuclear spins}

In almost all molecules there are distinct nuclei with various hyperfine couplings. In particular in toluene there are inequivalent $^1$H nuclei with either strong or weak hyperfine coupling \cite{wertz}. Since the ESR \cite{manassen} shows fairly strong hyperfine splitting of $\sim$10MHz, while sharp NMR lines result from  weakly coupled nuclei we are motivated to study the case of two nuclei spins, one spin with a strong hyperfine coupling, producing the dominant structure in the ESR, while the other with a weak coupling. In particular when the latter hyperfine coupling is weaker than the linewidth then it is not noticeable in the ESR. As we find here, the correlation of the weakly coupled nuclear spin produces a NMR, that is sharper as the ESR linewidth increases. 

We consider a single electron spin coupled to two nuclear spins. The Hamiltonian reads 
\beq{03} 
{\cal H}=\half\nu_e\sigma_z +\half \nu_{n,1} \tau_{z,1} + \half \nu_{n,2} \tau_{z,2} +\sigma_z\left(\tilde a_{1} \tilde \tau_{1} 
+\tilde a_{2} \tilde \tau_{2} \right) \ ,
\eeq
where ($j=1,2$)  
\be
\tilde \tau_{j} \equiv \frac{1}{\tilde a_{j}}\left(a_{j}\tau_{z,j} + d_{j}\tau_{y,j}\right)\quad,\quad \tilde a_{j} = \sqrt{a_{j}^{2}+d_{j}^{2}}\ .
\ee 

The Lindblad equation is identical to (\ref{eq:Lindblad}). We assume the nuclear spin $\vec \tau_{2}$ to have a strong 
hyper-fine coupling, thus providing the hyperfine-resolved spectrum. The nuclear spin $\vec \tau_{1}$ is weakly coupled. 
Therefore, as in the previous section, it is subject to motional narrowing and shows the NMR oscillations as well as the overlap with the ESR signal. 

\begin{figure}[b]
	\includegraphics*[height=.3\columnwidth]{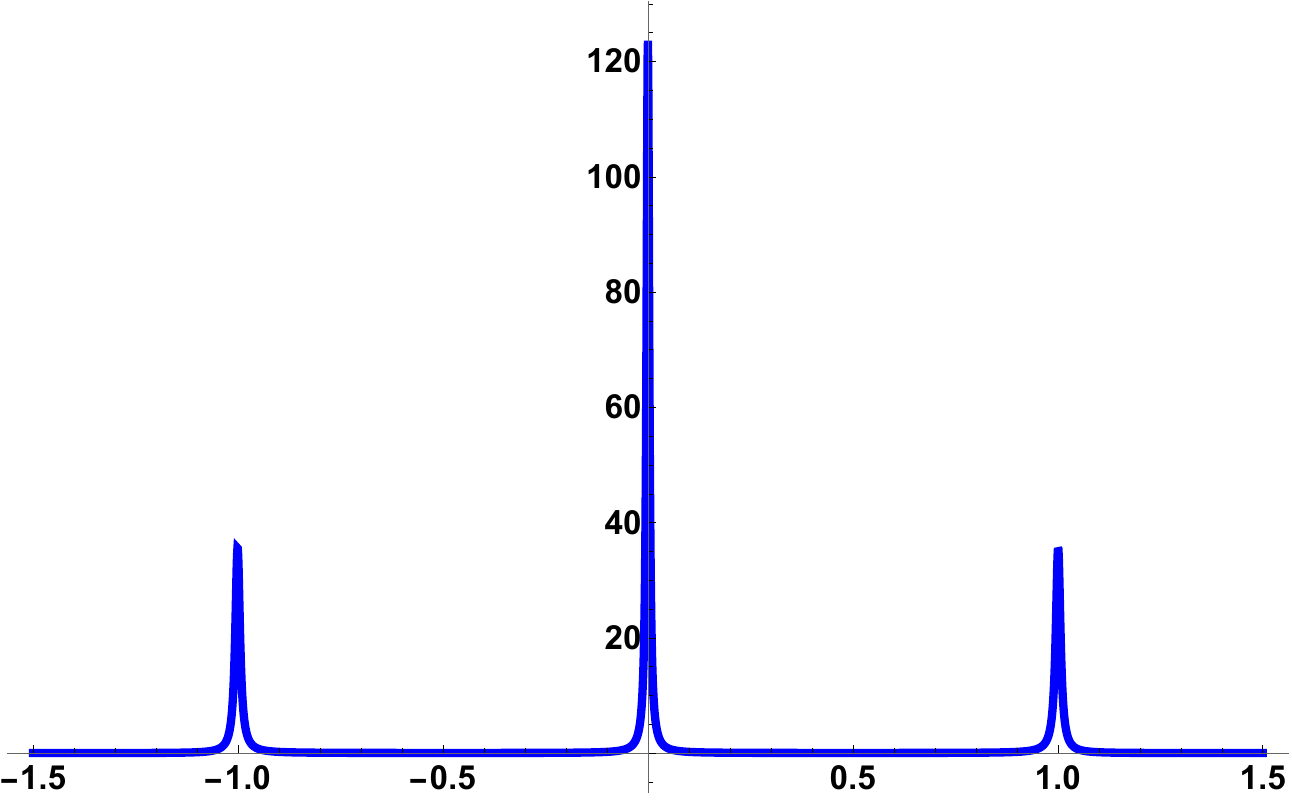}
	\includegraphics*[height=.3\columnwidth]{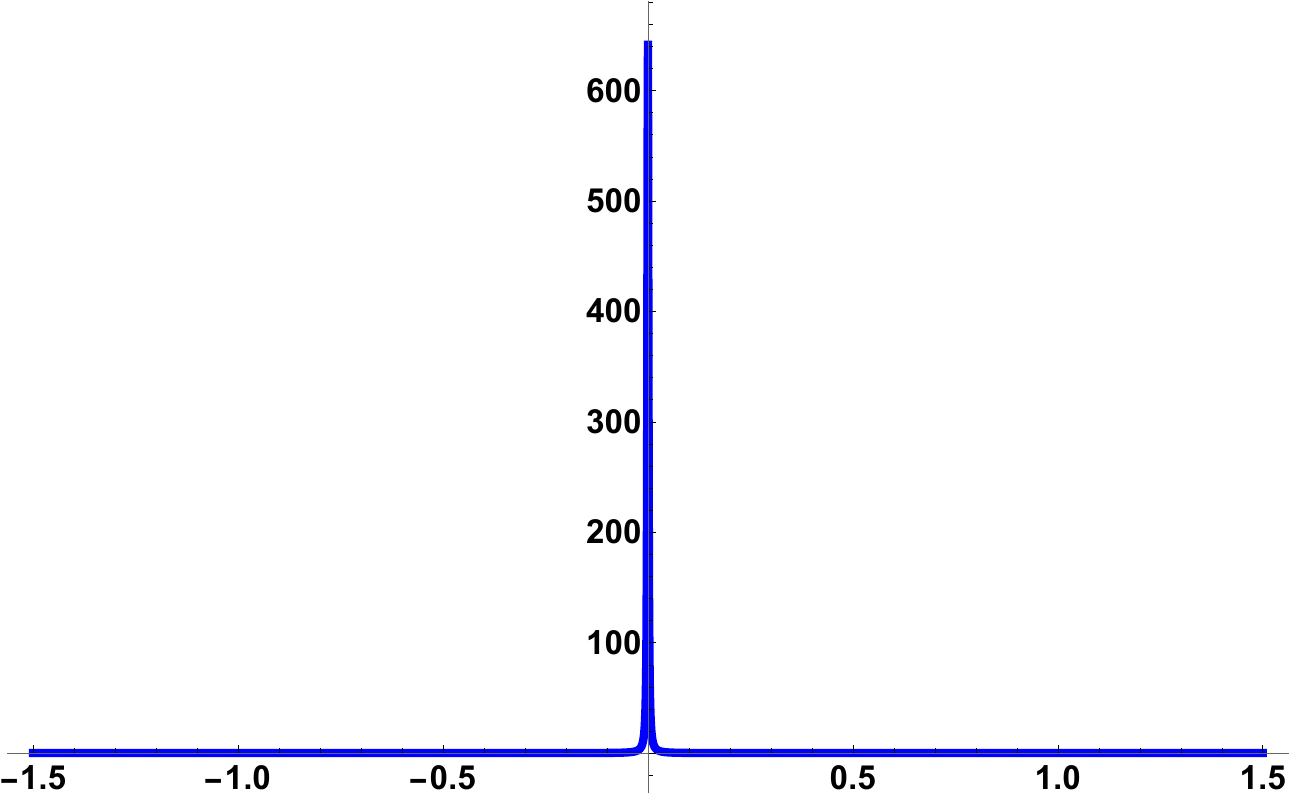}\\
	\includegraphics*[height=.28\columnwidth]{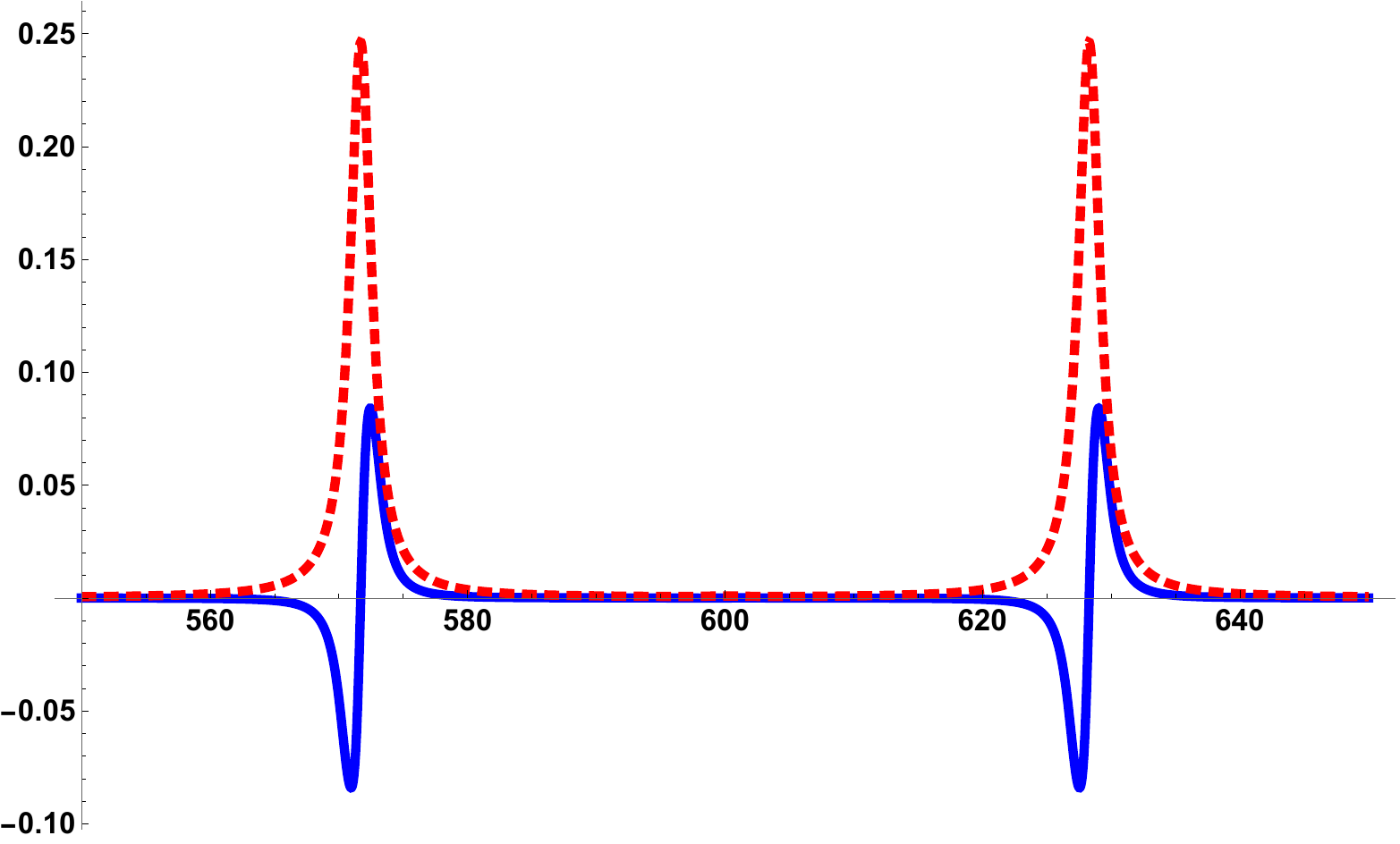}
	\includegraphics*[height=.28\columnwidth]{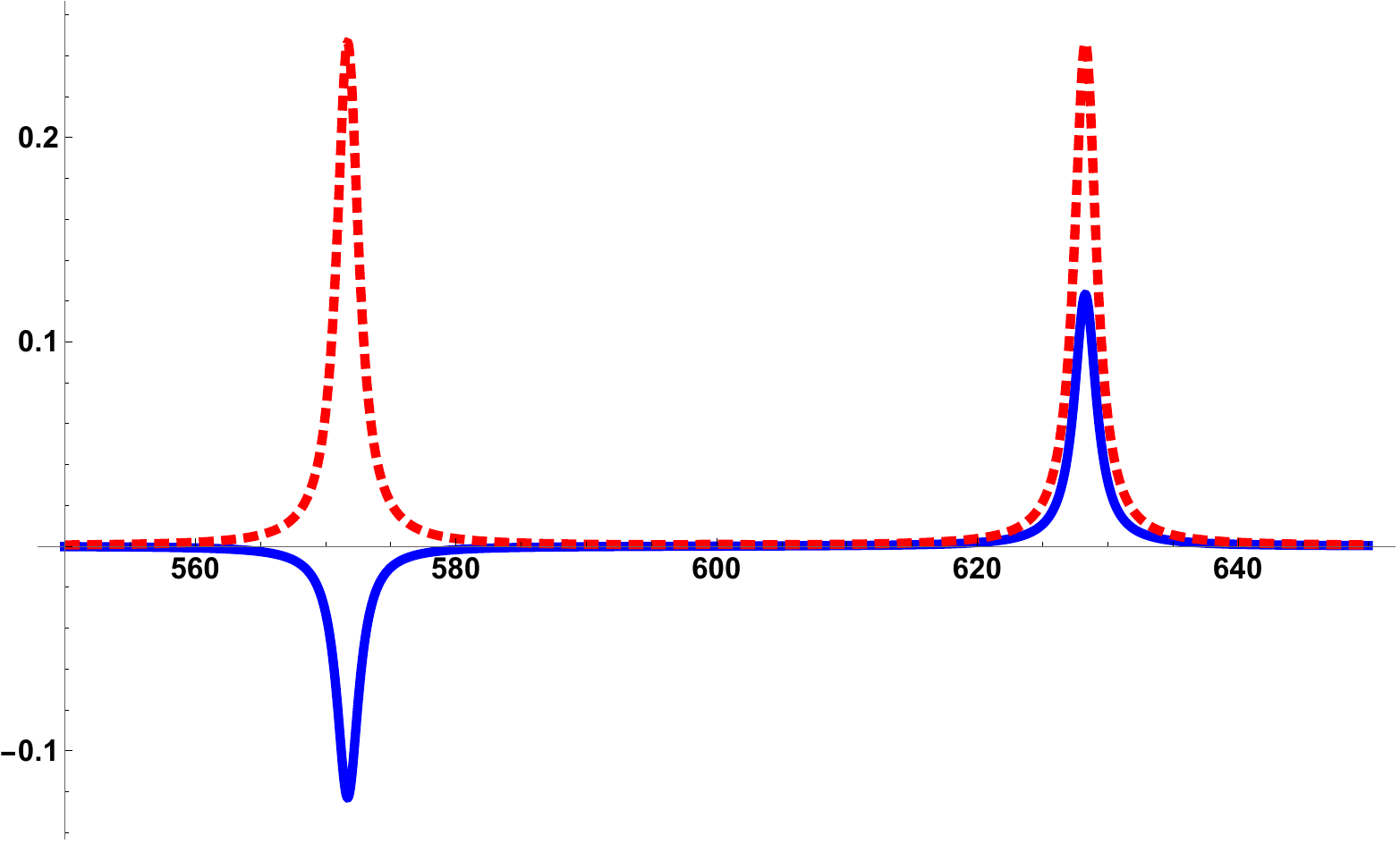}
	\caption{{\bf NMR with two nuclear spins. Left panel, upper part:} NMR Spectrum of the weakly coupled spin $\vec \tau_{1}$. 
		{\bf Left panel, lower part:} 
		The overlap function $\langle \hat C(\omega) \tilde \tau_{1} \rangle$ (blue, enhanced by factor 5 for clarity) and the ESR spectrum $C(\omega)$ (red, dashed) as functions of $\omega$.
		{\bf Right panel, upper part:} NMR Spectrum of the strongly coupled spin $\vec \tau_{2}$. This spin shows telegraph noise, but no coherent oscillations.  
		{\bf Right panel, lower part:} 
		The correlation function $\langle \hat C(\omega) \tilde \tau_{2} \rangle$ (blue, supressed by factor 2 for clarity) and the ESR spectrum $C(\omega)$ (red, dashed) as functions of $\omega$.
		Parameters are $\nu_e=600,\,\nu_{n,1}=1,\,\nu_{n,2}=1.1,\,a_{1}=0.05,\,d_{1}=0.05,\,a_{2}=10.0,\,d_{2}=10.0,\,\Gamma_{u}=\Gamma_{d}=1.0$.}
	\label{ESRoverlap2}
\end{figure}

The results are shown in Fig.~\ref{ESRoverlap2}.
We again observe correlations between the weakly coupled nuclear spin and the ESR signal. 
The latter is split by the strongly coupled nuclear spin.
We therefore consider this case with two nuclear spins as viable candidates accounting for the data \cite{manassen}.

\section{Two molecule case}

We assume here that in addition to the tested molecule that has a nuclear spin the system has an additional spectator molecule that can absorb an ionized electron from the tested one. The advantage of this model is that ionization is included and that non-radical and radical molecules are treated on equal footing. Radical molecules in equilibrium (e.g. TEMPO \cite{manassen}) become non-radicals upon ionization by an external bias, while non-radical molecules in equilibrium (e.g. Toluene \cite{manassen}) become radicals upon ionization. The case of two radical molecules is labelled as $|11\rangle$, it has two electron spins and one nuclear spin, hence a total of 8 states (assuming the nuclear spin is $\half$). The case of two non-radical molecules is labelled as $|20\rangle$, it has two states, i.e. the whole Hilbert space contains 10 states. 
Evidently, the measured ESR signal comes from the $|11\rangle$ state, however an NMR signal may come from either $|11\rangle$ or from $|20\rangle$. The situation is somewhat similar to that of two quantum dots \cite{petta} that by changing gate voltages can transfer charge between their $|20\rangle$ and $|11\rangle$ states.

The Hamiltonian in the two subspaces, respectively, as well as the total Hamiltonian are
\beq{06}
&&{\cal H}_{11}=\half\nu_1\sigma_z +\half\nu_2 s_z  +\half\nu_n\tau_z+a\sigma_z
\tau_z +d\sigma_z\tau_y \nonumber\\&&
{\cal H}_{20}=\half\nu_n\tau_z\nonumber\\&&
{\cal H}={\cal H}_{11}|11\rangle\langle 11|+{\cal H}_{20}|20\rangle\langle 20|
\eeq
where ${\bm \sigma},\,{\bf s},\,{\bm\tau}$ are Pauli matrices representing the two electron spins and the nuclear spin, respectively and $\nu_1,\,\nu_2$ are the electron resonance frequencies of the two radicals, respectively (in the absence of hyperfine couplings). An additional direct tunneling element between $|11\rangle$ and $|20\rangle$ is possible, as well as a chemical potential shift between these subspaces; we find that both terms have a minor effect on the results in this section.

We assume the switching between the $|11\rangle$ and $|20\rangle$ subspaces to be dissipative. We introduce the rate $\gamma_f$ for the $|20\rangle\rightarrow|11\rangle$ transition and $\gamma_d$ for the reverse process. Labeling the states with $\pm$ for $\sigma_z$, $s_z$ and $\tau_z$ spins, respectively, we order the $|11\rangle$ space as $|+++,++-,+-+,+--,-++,-+-,--+,---\rangle$. The jump operators couples $|20\rangle$ to just the singlet component of the $|11\rangle$ state, which is either $|0, 0, 0, -1, 0, 1, 0, 0 \rangle/\sqrt{2}$ or $|0, 0, -1, 0, 1, 0, 0, 0\rangle/\sqrt{2}$ (i.e. two nuclear spins). We then construct a $10\times 10$ operator jump J that transfers the $|20\rangle$ singlet states into $|11\rangle$, while its transpose $J^{T}$ is transferring the opposite way. The operator $J^T$, of dimension $10\times 10$, has 8 rows of zeroes while the 9th and 10th rows are
\beq{20}
\frac{1}{\sqrt{2}}\begin{pmatrix}0&0&-1&0&1&0&0&0&0&0 \\0&0&0&-1&0&1&0&0&0&0 \end{pmatrix} 
\eeq

If $\gamma_f\gg\gamma_b$ the dominant subspace is $|11\rangle$ representing an "ESR" state, while in the opposite case $|20\rangle$ is dominant, it has no ESR.

Since only singlet states are transferred by $\gamma_f,\,\gamma_b$ we consider additional spin relaxations within the $|11\rangle$ states. We define, as in Eq. \ref{eq:Lindblad}, spin flip rates for both molecules $\Gamma^{(i)}_d,\,\Gamma^{(i)}_u$, where $i=1,2$; in most examples below we consider the regime $\Gamma^{(i)}_d=\Gamma^{(i)}_u\equiv \Gamma^{(i)}$.
We define a $10\times 10$ operator $M_1=\sigma_+\otimes\mathbbm{1}\otimes\mathbbm{1}$ and 0 in  the $2\times 2$ space of $|2 0\rangle$, similarly $M_2=\mathbbm{1}\otimes s_+\otimes\mathbbm{1}$. Hence
the Lindblad equation is 
\beq{07}
\frac{d\rho}{dt}=L\cdot\rho=&&-i[{\cal H},\rho]+\gamma_f[J\rho J^\dagger-\half J^\dagger J\rho-\half \rho J^\dagger J]+
\gamma_b[J^\dagger\rho J-\half J J^\dagger\rho-\half \rho J J^\dagger] \nonumber\\&&
+\Gamma^{(1)}_{d}\left[M_1^\dagger\,\rho\,M_1 - \half M_1 M_1^\dagger\rho - \half \rho M_1 M_1^\dagger \right ] +
\Gamma^{(1)}_{u}\left[M_1\,\rho\,M_1^\dagger - \half M_1^\dagger M_1 \rho - \half \rho M_1^\dagger M_1 \right ]\nonumber\\&&
+\Gamma^{(2)}_{d}\left[M_2^\dagger\,\rho\,M_2 - \half M_2 M_2^\dagger\rho - \half \rho M_2 M_2^\dagger \right ] +
\Gamma^{(2)}_{u}\left[M_2\,\rho\,M_2^\dagger - \half M_2^\dagger M_2 \rho - \half \rho M_2^\dagger M_2 \right ]
\eeq

The supermatrices, as in Eq.  \eqref{super}, in a 10 dimensional Hilbert space are 
now  $100\times 100$. 
The steady state $\rho_\infty$ is then a 100 dimensional vector that solves $L\rho_\infty=0$.

\begin{figure}[t]
	\includegraphics*[height=.25\columnwidth]{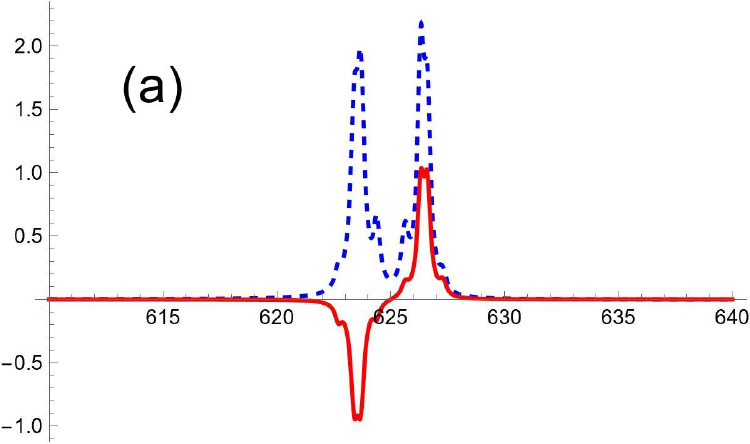}
	\includegraphics*[height=.25\columnwidth]{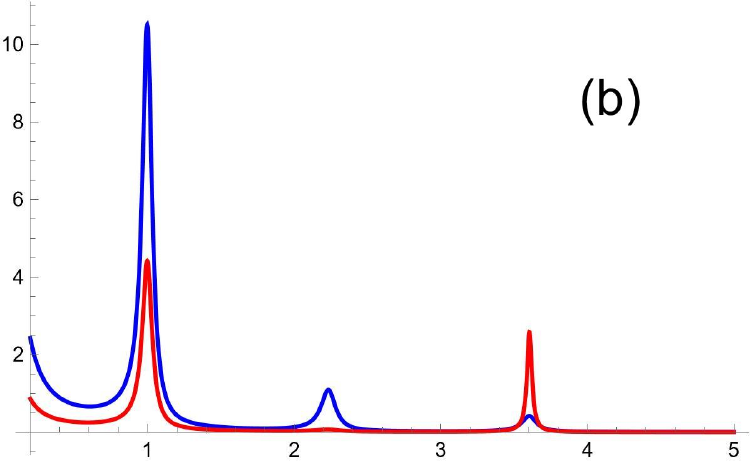}
	\caption{Two Molecules with strong intermolecular $\gamma_f=\gamma_b=10^4$ and and weak intramolecule relaxations. The electron Larmor frequencies are $\nu_1=600,\,\nu_2=650$ and the hyperfine couplings are $a=d=1$ while $\Gamma^{(1)}=\Gamma^{(2)}=.03$ (except the red curve in (b)). (a) ESR spectrum (blue dashed) and its overlap with the nuclear $\tilde\tau$ (red). (b) The nuclear correlation with no electron polarization, $\Gamma^{(1)}=\Gamma^{(2)}=.03$ (blue) and with polarization i.e.  $\Gamma^{(1)}_u=\Gamma^{(2)}_u=.05$ and $\Gamma^{(1)}_d=\Gamma^{(2)}_d=.01$ (red).}
	\label{2mol}
\end{figure}

We define an NMR operator for the two-molecule case that operates only in the $|11\rangle$ state where ESR is monitored,
\beq{19}
\bar \tau=\mathbbm{1}\otimes\mathbbm{1}\otimes\tilde\tau\cdot|11\rangle\langle 11|\,\oplus\, 0\cdot|20\rangle\langle 20|
\eeq
to be used instead of $\tilde\tau$ in the single molecule case. The exclusion of the $|20\rangle$ state reduces the amplitude 
of the observed NMR signal. The exclusion of $|20\rangle$ is essential since in the expansion of $\hat C(\omega)$ Eq. \eqref{e71} only the $|11\rangle$ component affects the measure ESR.

We have found new scenarios for NMR when allowing for large relaxation rates $\gamma_f,\,\gamma_b$, though still small on the voltage scale (the chemical potential difference of the two molecules is $\sim 1$V).  We note that in this case the singlet states rapidly decay while the $\Gamma^{(1)},\,\Gamma^{(2)}$ transitions populate the triplet states, leading to dominance of the latter, hence a strong coupling between the two spins. Projecting the Hamiltonian (keeping only the $a,\, d$ hyperfine terms) onto the triplet states we find that the ESR frequencies are
\beq{21}
\nu_\pm^{ESR}=\half[\nu_1+\nu_2\pm\nu_n\pm\sqrt{4(a^2+d^2)-4a\nu_n+\nu_n^2}]
\eeq
These frequencies can be detected by measuring ESR of either spin and provide an experimental tool for identifying $\gamma_f,\,\gamma_b$, i.e. the ESR lines would shift towards these frequencies as $\gamma_f,\,\gamma_b$ increase.

We consider first cases with weak ESR linewidths $\Gamma^{(1)},\,\Gamma^{(2)}$. The ESR spectrum is shown in Fig. \ref{2mol}a, showing large overlap/esr ratio $\sim O(1)$
and sharp NMR frequencies at at
$\tilde\nu_n=\sqrt{(\nu_n\pm 2a)^2+(2d)^2}$ as in the single molecule case  (Fig. \ref{ESRoverlapresolved}), however we find in this 2-molecule case an NMR signal also at the bare $\nu_n$. We note that these frequencies do not shift with an electron polarization since for weak $\Gamma^{(1)},\,\Gamma^{(2)}$ they correspond to $\sigma_z=\pm$, independent of spin polarization. In fact these cases correspond to extreme chemical shifts, hence do not depend on shifts in $\langle \sigma_z\rangle \neq \pm 1$. The intensities of these lines, however, do depend on the electron polarization, see Fig. \ref{2mol}b.
We note that this scenario with $\Gamma^{(1)},\Gamma^{(2)}<\nu_n$ is not realizable in common molecules, unlike the following case.

\begin{figure}[b]
	\includegraphics*[height=.3\columnwidth]{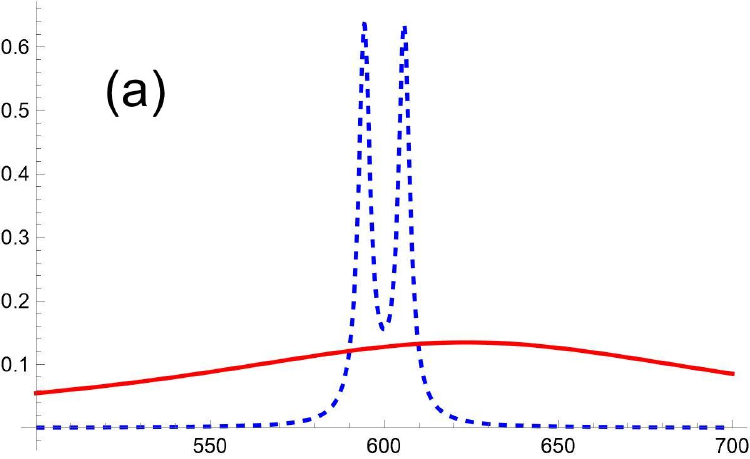}
	\includegraphics*[height=.3\columnwidth]{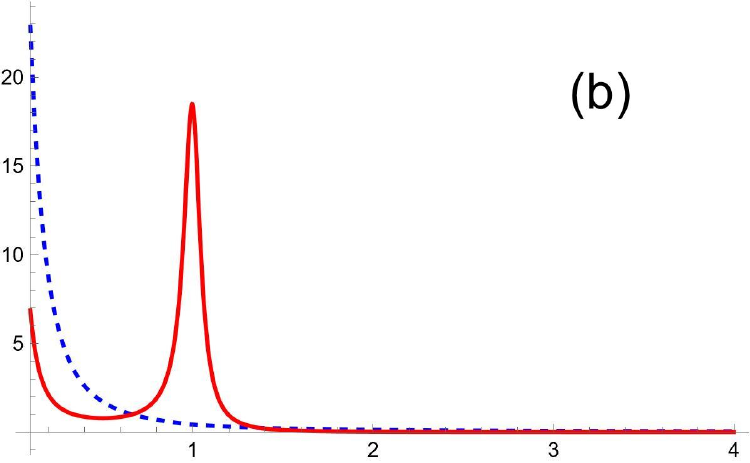}
	\caption{{\bf Effect of intermolecular transition}: $\nu_1=600,\,\nu_2=650,\,\nu_n=1,\Gamma^{(1)}=2,\,\Gamma^{(2)}=200,\,a=d=2$. Compare the cases of strong $\gamma_f=\gamma_b=10^4$ (red curves) and weak $\gamma_f=\gamma_b=0.1$ (blue dashed curves) for (a) ESR spectrum and (b) NMR spectrum.}
	\label{intermolecular}
\end{figure}

A second scenario that shows an NMR signal occurs when the coupled molecules are very different. Consider the first molecule with parameters that would show hyperfine split ESR, e.g. $\Gamma^{(1)}=2$ and $a=d=2$, however it does not show NMR since the condition $\tilde a^2/\Gamma^{(1)}\ll\nu_n$ is not satisfied. The second molecule has a large $\Gamma^{(2)}=200$, which has no effect if the intermolecular coupling is weak ($\gamma_f=\gamma_b=0.1$), see blue lines in Fig. \ref{intermolecular}. However, when the intermolecular coupling becomes large, e.g. $\gamma_f=\gamma_b=10^4$, the ESR spectrum corresponds to a triplet whose linewidth is dominated by the large $\Gamma^{(2)}$. In this case the ESR does not show hyperfine peaks, however the NMR is seen, see the red lines in Fig. {\ref{intermolecular}. This figure shows the trade-off between having a small or large electron relaxation: weak relaxation shows ESR with hyperfine split lines but no NMR while strong relaxation makes the ESR featureless while an NMR signal is present. This NMR line shifts with the electron polarization, as realized by $\Gamma^{(2)}_u\neq\Gamma^{(2)}_d$ of the stronger relaxation. We infer then that the NMR signal is from the $|11\rangle$ subspace.
	
	The scenario of Fig. \ref{intermolecular} can be of practical use. Assume the common situation that we have a molecule with weak $\Gamma^{(1)}$ that shows hyperfine lines, yet it does not show NMR (blue lines). Now imagine that we have at our disposal a second molecule with large electron relaxation $\Gamma^{(2)}$, a molecule  that we can move to the vicinity of the first molecule, generating significant couplings $\gamma_f,\gamma_b$. This procedure generates detection of the NMR signal of the first molecule, see red lines in Fig. \ref{intermolecular}.

	\section{Non-equilibrium}
	
	We consider here the situation \cite{manassen} where the STM voltage is modulated periodically between a strong and weak value, as often done in the experiment \cite{manassen}. The Hamiltonian is as defined in \eqref{e06} and 
	\beq{100}
	{\cal H}_{1,2}={\cal H}_{11}|11\rangle\langle 11|
	+({\cal H}_{20}\pm V)|20\rangle\langle 20|
	\eeq
	where ${\cal H}_1$ corresponds to $+V$ and acts during a time $T_1$ while ${\cal H}_2$ corresponds to $-V$ and acts during $T_2$. The voltage alternation is then repeated with a total periodicity of $\bar T=T_1+T_2$. We assume that upon voltage change the relaxation rates also switch $\gamma_f\leftrightarrow \gamma_b$, representing stronger relaxation towards the state with the lower voltage.
	This formalism is fully equivalent to the equilibrium case when $V=0$ and without $\gamma_f\leftrightarrow\gamma_b$ switching, e.g. when $\gamma_f=\gamma_b$.
	
	The Lindblad operators, exhibited now as super-operators, are then
	\beq{09}
	LL_1=&&-i{\cal H}_1\otimes I_d+iI_d\otimes{\cal H}_1^T+\half\gamma_f (2J\otimes (J^\dagger)^T-J^\dagger\cdot J\otimes I_d-I_d\otimes[J^\dagger\cdot J]^T)\nonumber\\&& +\half\gamma_b (2J^\dagger\otimes J^T-J\cdot J^\dagger\otimes I_d-I_d\otimes[J\cdot J^\dagger]^T)\nonumber\\ LL_2=&&-i{\cal H}_2\otimes I_d+iI_d\otimes{\cal H}_2^T+\half\gamma_b (2J\otimes (J^\dagger)^T-J^\dagger\cdot J\otimes I_d-I_d\otimes[J^\dagger\cdot J]^T)\nonumber\\&& +\half\gamma_f (J^\dagger\otimes J^T-J\cdot J^\dagger\otimes I_d-I_d\otimes[J\cdot J^\dagger]^T)\nonumber\\
	MM_1=&&\half\Gamma^{(1)}_d[2M_1^\dagger\otimes M_1^T-M_1\cdot M_1^\dagger\otimes I_d-I_d\otimes (M_1\cdot M_1^\dagger)^T]\nonumber\\&&
	\half\Gamma^{(1)}_u[2M_1\otimes (M_1^\dagger )^T-M_1^\dagger\cdot M_1\otimes I_d-I_d\otimes (M_1^\dagger\cdot M_1)^T]
	\eeq
	and similarly for $MM_2$ with $1\rightarrow 2$. The evolution operators for the two time periods are then
	\beq{10}
	&&L_1=LL_1+MM_1+MM_2,\qquad \frac{d\rho}{dt}=L_1\rho\Rightarrow\qquad \rho(T_1)=\eexp{L_1T_1}\rho(0)\nonumber\\&&
	L_2=LL_2+MM_1+MM_2,\qquad \frac{d\rho}{dt}=L_2\rho\Rightarrow \rho(T_1+T_2)=\eexp{L_2T_2}\eexp{L_1T_1}\rho(0)
	\eeq
	Hence the evolution operator for the time $\bar T$ is
	$U=\eexp{L_2T_2}\eexp{L_1T_1}$. 
	In the simulations below we use $T_1=T_1=.05\,\mu$sec, we find that the results are insesitive to the values of $T_1,\,T_2$ as long as they are in the window $1/\nu_1\ll T_1+T_2\ll 1/\nu_n$.
	
	Define eigenvectors $U\rho_i=\lambda_i\rho_i$, hence an expansion $\rho(0)=\sum_{n=1}^{100}c_n\rho_n$ yields after $N$ steps, each of duration $\bar T=T_1+T_2$,
	\beq{11}
	\rho(N(\bar T))=\sum_n c_n\lambda_n^N\rho_n\,.
	\eeq
	We expect that only $n=1$ has $\lambda_1=1$ and $\tr[\rho_1]=1$, i.e. $\rho_1$ is the steady state while for $n>1$
	$|\lambda_n|<1$ and $\tr[\rho_n]=0$ (otherwise the steady state is not unique).
	The eigenvalues for $n>1$ $\lambda_n=|\lambda_n|\eexp{i\varphi_n}$ correspond to eigenfrequencies $\nu_n^*=\varphi_n/\bar T$ with linewidth $\Gamma_n=-(\ln \lambda_n|)/\bar T$. One of the $\nu_n^*$ will be identified with the nuclear frequency.
	
	For the numerical program (using Mathematica), it is more efficient to find the eigenvectors $v^{(1)}_i$ and eigenvalues $e^{(1)}_i$ of $L_1$ and then build a matrix $E_1$ of size $100\times 100$ whose columns are $e^{(1)}_i$. In terms of a diagonal matrix $D_1$ whose elements are $\eexp{e^{(1)}_i T_1}$ the evolution during $T_1$ is
	\beq{12}
	U_1=E_1D_1(E_1)^{-1}
	\eeq
	and similarly with $1\rightarrow 2$. 
	Note that the eigenvectors are not orthogonal ($L_1$ is not hermitian), we need, however, that $E_1$ is invertable, otherwise the set of $\rho_i$ cannot span the whole space. The parameter subspace where this condition is violated is known as exceptional points \cite{bergholtz} and is of measure zero.
	
	We consider the general case $L_1\neq L_2$ and evaluate the nuclear correlation using $\bar\tau$ of Eq. \eqref{e19}. In the supermatrix notation the density matrices $\rho_n$ in Eq. \eqref{e11} are vectors (of length 100). Assuming that these vectors span the whole space then $\bar\tau\cdot\rho_k=\sum_n d_{kn}\rho_n$. The correlation function, using the regression theorem for $N_1>N_2$ where $N_i$ correspond to times $N_i \bar T$, is
	\beq{15}
	&&C_{nmr}(N_1-N_2)=\langle \bar\tau(N_1)\bar\tau(N_2)\rangle=\tr[\bar\tau U^{N_1-N_2}\bar\tau\rho_1]=
	\sum_n d_{1n}\lambda_n^{N_1-N_2}\tr[\bar\tau\rho_n]=\sum_n d_{1n}d_{n1}\lambda_n^{N_1-N_2}\nonumber\\
	C_{nmr}(\nu)&&=\sum_n d_{1n}d_{n1}\int_0^\infty \eexp{-\Gamma_n t-i\nu_n^* t +i\nu t}dt+c.c.=\sum_n d_{1n}d_{n1}\frac{1}{-i(\nu-\nu_n^*)+\Gamma_n}+c.c.
	\eeq
	where $\tr[\rho_n]=\delta_{n,0}$ and the c.c. comes from summation on $N_1<N_2$.
	We assume here that the probed frequency is small, i.e. $\nu_n^*\bar T\ll 2\pi$ so that the discrete Fourier transform with respect to $N$ can be replaced by integration, i.e. $N\rightarrow t/\bar T$. Using a matrix $E$ of eigenvectors $E_{ij}=(\rho_j)_i$ we obtain for the matrix $\hat d$, $(\hat d)_{kn}=d_{kn}$
	\beq{16}
	(\tilde\tau\cdot E)_{ik}&&=\sum_j(\tilde\tau)_{ij}(\rho_k)_j=\sum_n d_{kn}(\rho_n)_i=\sum_n d_{kn}E_{in}
	=(E\cdot\hat d^{T})_{ik}\nonumber\\
	\Rightarrow\qquad \hat d&&=(E^{-1}\cdot \tilde\tau\cdot E)^{T}
	\eeq
	which is an efficient way of evaluating $d_{kn}$. Thus finally
	\beq{20}
	C_{nmr}(\nu)=\sum_{n>0}\re[ \frac{2d_{1n}d_{n1}}{-i(\nu-\nu_n^*)+\Gamma_n}]
	\label{nonequiNMR}
	\eeq
	
	\begin{figure}[tb]
		\includegraphics*[height=.25\columnwidth]{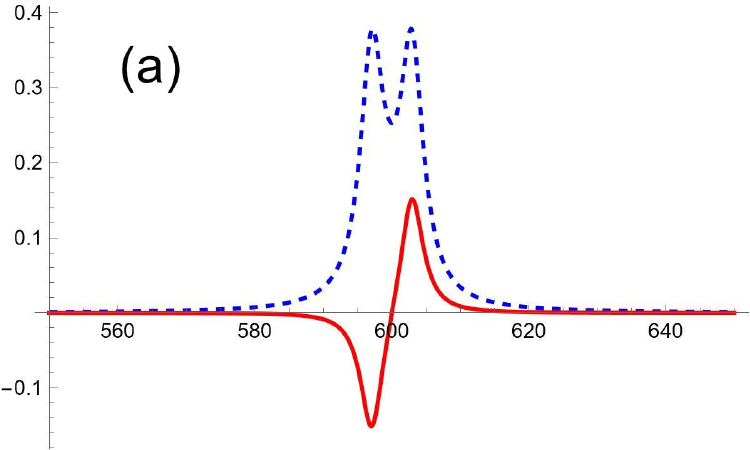}
		\includegraphics*[height=.25\columnwidth]{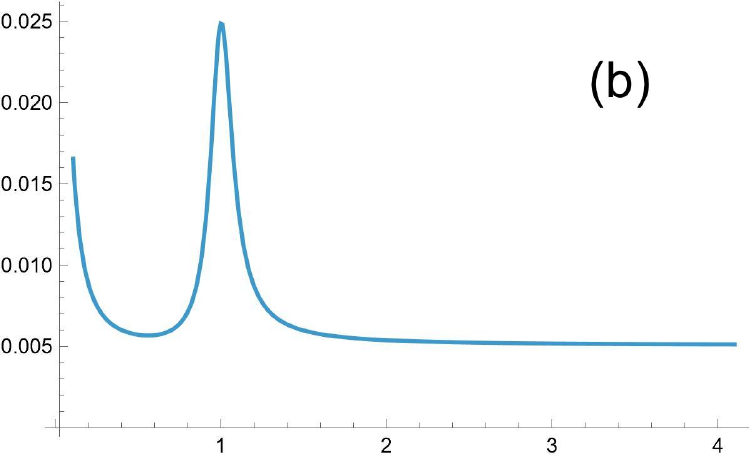}
		\caption{{\bf Non-equilibrium.} Consider two molecules with $\gamma_f=10^4,\,\gamma_b=0.1,\,V=0$ and  $\nu_1=600,\,\nu_2=650,\,\nu_n=1,\Gamma^{(1)}=2,\,\Gamma^{(2)}=200,\,a=d=1$. (a) ESR and overlap using equilibrium expressions Eqs. (\ref{C0},\ref{nuclearESR}) averaging on the periods $T_1,\,T_2$. (b) NMR spetra using the non-equilibrium expression \eqref{nonequiNMR}.}  
		\label{nonequi}
	\end{figure}
	
	We find numerically that allowing $V\neq 0$ has a minor effect on the previous results. However, allowing $\gamma_f\neq\gamma_b$ leads to a significant effect if $\gamma_f/\gamma_b\gg 1$, as shown in Fig. \ref{nonequi}. The case $\gamma_f/\gamma_b\gg 1$ means that the relaxation $|20\rangle\rightarrow |11\rangle$ is dominant so that in the period $T_1$ the dominant subspace is $|11\rangle$, while $|20\rangle$ is dominant in the period $T_2$. The $|11\rangle$ component allows detection of ESR with hyperfine splitting, Fig. \ref{nonequi}a. The ESR is evaluated using the equilibrium formulation Eq. \ref{C0} with either $L\rightarrow L_1$ or $L\rightarrow L_2$, since there are many ESR oscillations in one period $T_1$ or $T_2$; the actual measurement would average the $T_1$ and $T_2$ data (with $T_1$ dominant), shown in Fig. \ref{nonequi}a. We also evaluate the NMR spectrum, Fig. \ref{nonequi}b, using the non-equilibrium expression Eq. (30), showing a clear signal. 
	
	\section{Conclusions}
	
	We have found a number of scenarios where real time analysis of ESR-STM data shows NMR signals.  All these scenarios are summarized in table I, for easy access. Note that in all the scenarios we find that an off-diagonal hyperfine coupling $\sim \sigma_z\tau_y$ is essential for NMR observation. In appendix A we show that a rotated molecule, i.e. the magnetic field is not parallel to a principal axis of the molecule, has in general such an off-diagonal coupling.

	We have shown that sharp NMR lines are possible with either $\Gamma_1\ll\nu_n$ or with $\frac{\tilde a^2}{\Gamma_1}\ll \nu_n$, as in the simplified Eq. \eqref{gaussian} and in appendix B.
	The case with $\Gamma_1\ll\nu_n$, Figs. 2 and 4 show NMR lines at shifted positions and when two molecules are coupled also at the bare nuclear frequency. This situation of very weak electron relaxation, however, does not apply to known molecules. 
	
	We therefore focus on the case $\frac{\tilde a^2}{\Gamma_1}\ll \nu_n$, a motional narrowing phenomenon. For a single molecule we consider two nuclei, the dominant one with a strong $\tilde a$ that allows observable ESR with hyperfine splitting, while the second nuclus has weak $\tilde a$ that allows for a sharp NMR. This scenario applies e.g. to toluene that has many $^1$H nuclei with varying hyperfine couplings.
	
	For cases with a single nucleus in the molecule, e.g. $^{31}$P or $^{14}$N, we propose scenarios with strongly coupled two molecules such that the spectator molecule has a much larger relaxation rate. Fig. \ref{intermolecular} shows that NMR appears at strong $\gamma_f,\,\gamma_b$ at the expense of hyperfine structure in ESR. The ESR data on cases with sharp NMR are partial \cite{manassen}, one needs more detailed data on a wider frequency range to determine if this scenario applies.
	In fact, if one can move a spectator molecule with strong relaxation to the vicinity of a target molecule, than any target nuclear frequency can be detected.
	
	Our final scenario is a non-equilibrium one, i.e. two molecules with different intra-molecular relaxation rates, as above, and also $\gamma_f/\gamma_b\gg 1$. As shown in Fig. \ref{nonequi} one can have both ESR with hyperfine structure and a sharp NMR. The relaxation rates are controlled by the STM voltage. The intermolecular relaxation can be deduced from detailed ESR spectra, once available.

		\begin{table}[t]
		\caption{ Parameter regimes for observing sharp NMR lines in terms of the ESR linewidth $\approx\Gamma_1$ (or $\Gamma^{(1)},\,\Gamma^{(2)}$ for two molecules), hyperfine coupling $\tilde a$ defined in Eq. \eqref{e02} (or $\tilde a_1,\,\tilde a_2$ for two nuclei) and transition rates $\gamma_f,\,\gamma_b$ for charge transfer between two molecules. All cases have a single nucleus with Larmor frequency $\nu_n$, except when two nuclei are indicated, with comparable $\nu_n$. Also shown is the representative figure number.}
		\def\arraystretch{2.3}
			\begin{tabular}{|c|c|c|c|} Configuration & Parameter regimes & ESR shape & Figure \\ \hline
				single molecule & $\frac{\tilde a^2}{\Gamma_1}\ll \nu_n$ 
				& wide resonance  & 1 \\
				single molecule & $\Gamma_1\ll \nu_n,\,\tilde a$ & $\tilde a$ splitting &  2 \\
				single molecule, two nuclei &  $\frac{\tilde a_1^2}{\Gamma_1}\ll \nu_n,\,\, \tilde a_2>\Gamma_1$ & $\tilde a_2$ splitting & 3\\
				two molecules & $\Gamma^{(1)},\,\Gamma^{(2)}\ll \nu_n,\,\tilde a \qquad \gamma_f,\,\gamma_b \gg \tilde a$ & triplet, $\tilde a$ splitting & 4 \\
				two molecules & $\frac{\tilde a^2}{\Gamma^{(2)}}\ll \nu_n,\,\Gamma^{(1)}<\tilde a \qquad \gamma_f,\gamma_b\gg \tilde a $ & wide resonance & 5 \\
				two molecules, $\gamma_f, \gamma_b$ alternating & $\frac{\tilde a^2}{\Gamma^{(2)}}\ll \nu_n,\,\Gamma^{(1)}<\tilde a \qquad \gamma_f\gg \tilde a > \gamma_b$ & $\tilde a$ splitting & 6 \\
				\hline
			\end{tabular}
		\label{table1}
	\end{table}

	\acknowledgements
	We thank Yishay Manassen for valuable  discussions and insight into his data. This work was supported by Horizon-EIC-2022-Pathfinderopen-01 on 4D-NMR grant agreement no. 101099676.
	\appendix
	
	\section{Rotated hyperfine}
	Suppose that the hyperfine tensor has principal axes $z', y', x' $ at some orientation relative to the magnetic field, chosen in the $z$ direction. The diagonal hyperfine elements are $c',b',a'$. We wish to rotate this tensor and find the dominant hyperfine splitting for electron Larmor frequency $\nu\gg a',b',c'$.  Assume for simplicity that it is sufficient to rotate around the $x=x'$ axis with an angle $\theta$, this is the case if $c'$ is relatively small or if two elements are equal $b'=c'$. The hyperfine coupling with the rotation matrix $R(\theta)$ becomes, where the electron spin is $\bf S$, the nuclear spin is $\bf I$ and primes denote rotated quantities,
	\beq{04}
	S'\cdot A'\cdot I'=&&(S_z,S_y,S_x)\left(\begin{array}{ccc} \cos\theta &\sin\theta & 0 \\ -\sin\theta & \cos\theta & 0 \\
		0 & 0& 1 \end{array}\right)\left(\begin{array}{ccc} a' & 0 &0 \\0 & b' & 0 \\ 0 & 0& c'\end{array}\right)
	\left(\begin{array}{ccc} \cos\theta &-\sin\theta & 0 \\ \sin\theta & \cos\theta & 0 \\
		0 & 0& 1 \end{array}\right)\left(\begin{array}{c} I_z \\ I_y \\ I_x \end{array}\right)\nonumber\\&&
	=(S_z,S_y,S_x)\left(\begin{array}{ccc} a'\cos^2\theta+b'\sin^2\theta & \sin\theta\cos\theta(b'-a') & 0 \\
		\sin\theta\cos\theta(b'-a') & a'\cos^2\theta+b'\sin^2\theta & 0 \\ 0 & 0 & c' \end{array}\right)
	\left(\begin{array}{c} I_z \\ I_y \\ I_x \end{array}\right)
	\eeq
	where the rotated spin vectors ${\bf S}={\bf S}'\cdot R(-\theta),\, {\bf I}=R(\theta)\cdot {\bf I}'$. For a general rotation all elements of the tensor become finite. In presence of a Larmor frequency $\nu$ in the Hamiltonian, i.e. $ \nu S_z$, the $S_x,\,S_y$ terms are perturbative of order (hyperfine)$^2/\nu$, hence the dominant hyperfine term is
	\beq{05}
	&&{\cal H}_{hyp}=S_z\,(aI_z+dI_y)=\sqrt{a^2+d^2} S_z\, \tilde I, \qquad
	\tilde I=\frac{a}{\sqrt{a^2+d^2}}I_z+\frac{d}{\sqrt{a^2+d^2}}I_y\\&&
	a=a'\cos^2\theta+b'\sin^2\theta,\qquad d=\sin\theta\cos\theta(b'-a'),\qquad \Rightarrow a^2+d^2=a'^2\cos^2\theta+b'^2\sin^2\theta\nonumber
	\eeq
	The eigenvalues $\pm\sqrt{a'^2\cos^2\theta+b'^2\sin^2\theta}$ are in agreement with Eq. 1.66 (or  Eq. 3.44) of Ref. \onlinecite{abragam} (the g factor is assumed isotropic).  The anisotropic hyperfine tensor with off-diagonal elements is also studied in sections 7-6, 7-7 of Ref. \onlinecite{wertz}, including analysis of experimental data.

	\section{Nuclear dephasing for $\nu_n=0$}
	
	Consider NMR in a single radical molecule, i.e. one electron and one nuclear spin. 
	Consider the Hamiltonian Eq. \eqref{single}, 
	\beq{24}
	{\cal H}=
	\half\nu\sigma_z+\half\nu_n\tau_z+\tilde a\sigma_z\tilde\tau
	\eeq
	In addition there are electron relaxation rates $\Gamma_d,\,\Gamma_u$. The case of strong relaxations is discussed in Eq. \eqref{gaussian}. We extend this derivation to a general $\Gamma$ where $\Gamma\equiv \Gamma_d=\Gamma_u$, and as in \eqref{gaussian} with $\nu_n$ neglected.
	We use following trick \cite{schriefl}:
	Define conditional averages $\chi_\pm(t)=\langle \eexp{i2\tilde a\int^t \sigma_z(t')dt'}\rangle$, averaged over switching histories ending at $\sigma_z(t)=\pm 1$. These satisfy the rate equations
	\beq{26}
	&&\dot\chi_+=2i\tilde a\chi_+-\Gamma\chi_++\Gamma\chi_- ,\qquad
	\dot\chi_-=-2i\tilde a\chi_--\Gamma\chi_-+\Gamma\chi_+\nonumber\\&&
	\chi_{1,2}=\chi_+\pm\chi_-\,\Rightarrow\qquad\dot\chi_1=2i\tilde a\chi_2,\qquad \dot\chi_2=2i\tilde a\chi_1-2\Gamma\chi_2\nonumber\\&&
	\Rightarrow \ddot\chi_2=-4\tilde a^2\chi_2-2\Gamma\dot\chi_2\,\Rightarrow \qquad \omega_\pm=i\Gamma\pm\sqrt{4\tilde a^2-\Gamma^2}
	\eeq
	There are two solutions $\sim\eexp{i\omega_\pm t}$, with the initial conditions $\chi_\pm(0)=\half$ we have
	\beq{27}
	&&\chi_2(t)=\frac{\tilde a}{\sqrt{4\tilde a^2-\Gamma^2}}\eexp{-\Gamma t}[\eexp{i\sqrt{4\tilde a^2-\Gamma^2}\,t}-
	\eexp{-i\sqrt{4\tilde a^2-\Gamma^2}\,t}]\nonumber\\&&
	\chi_1(t)=\eexp{-\Gamma t}[\cos(\sqrt{4\tilde a^2-\Gamma^2}\,t)+\frac{\Gamma}{\sqrt{4\tilde a^2-\Gamma^2}}
	\sin(\sqrt{4\tilde a^2-\Gamma^2}\,t)]
	\eeq
	Our needed average is $\chi_1(t)$, it defines a dephasing rate via $\chi_1(t)\sim \eexp{-\gamma_\varphi t}$. Hence for $\Gamma< 2\tilde a$ we have $\gamma_\varphi=\Gamma$ apart for oscillating terms. For $\Gamma> 2\tilde a$ at long times $\chi_1(t)\sim \half\eexp{-\Gamma t+\sqrt{\Gamma^2-4\tilde a^2}\,t}$, hence the dephasing rate is
	\beq{28}
	&&\gamma_\varphi=\Gamma-\sqrt{\Gamma^2-4\tilde a^2}\qquad\qquad \Gamma>2\tilde a\nonumber\\&&
	\gamma_\varphi\rightarrow\frac{2\tilde a^2}{\Gamma}\,\,\,\qquad\qquad\qquad\qquad \Gamma\gg 2\tilde a
	\nonumber\\&&
	\gamma_\varphi=\Gamma\qquad \qquad\qquad\qquad\qquad  \Gamma<2\tilde a
	\eeq
	The $\Gamma\gg 2\tilde a$ case agrees with Eq. \eqref{gaussian}, useful for interpretation of many of our scenarios. One can extend this derivation to the case $\Gamma_d\neq \Gamma_u$ and derive, to leading order in $(\tilde a/\Gamma_1)^2$ the result given in \eqref{gaussian}.
	The case $\Gamma<2\tilde a$ is less relevant to experimental data, however it is complementary and motivates our small $\Gamma$ examples in Figs. 2 and 4.

\end{document}